# Predicted novel insulating electride compound between alkali metals lithium and sodium under high pressure[*]

Yang-Mei Chen (陈杨梅) [1, 2], Hua Y. Geng (耿华运) [2, †], Xiao-Zhen Yan (颜小珍) [3], Zi-Wei Wang (王紫薇) [2], Xiang-Rong Chen (陈向荣)[1, †], and Qiang Wu (吴强)[2, †]

[1]*Institute of Atomic and Molecular Physics, College of Physical Science and Technology, Sichuan University, Chengdu 610065, China*
[2]*National Key Laboratory of Shock Wave and Detonation Physics, Institute of Fluid Physics, CAEP; P.O. Box 919-102, Mianyang 621900, China*
[3]*School of Science, Jiangxi University of Science and Technology, Ganzhou 341000, China*

The application of high pressure can fundamentally modify the crystalline and electronic structures of elements as well as their chemical reactivity, which could lead to the formation of novel materials. Here, we explore the reactivity of lithium with sodium under high pressure, using a swarm structure searching techniques combined with first-principles calculations, which identify a thermodynamically stable LiNa compound adopting an orthorhombic oP8 phase at pressure above 355 GPa. The formation of LiNa may be a consequence of strong concentration of electrons transfer from the lithium and the sodium atoms into the interstitial sites, which also leads to opening a relatively wide band gap for LiNa-op8. This is substantially different from the picture that share or exchange electrons in common compounds and alloys. In addition, lattice-dynamic calculations indicate that LiNa-op8 remains dynamically stable when pressure decompresses down to 70 GPa.

## 1. Introduction

Alkali metals are widely studied for the primary understanding of the physics of interelectronic interactions between simple s electrons in the field of simple geometric ionic lattices. At ambient conditions, all the alkali metals crystallize in the bcc structure[1, 2] and display a free-electron like metallic character[3, 4]. Application of pressure on these systems results in more complex structures[5, 6] and remarkable physical phenomena such as unusual melting behavior[7, 8], Fermi-surface nesting[9], phonon instabilities[10], and superconductivities[11-14], and transformations into poor metals or even insulators[15-19].

At ambient conditions, the ionic radius of Li has large disparity with respect to the other alkali metals[20]. Therefore, only Li interalkalies (Li-Na, Li-K, Li-Rb, and Li-Cs) exhibit phase-separation behavior between different alkali metal elements[21]. The formation enthalpies of these Li interalkalies were calculated to be positive and rapidly increased with the size mismatch, wherein the Li-Cs system has the highest positive formation enthalpies and the most immiscibility. Pressure, as an efficient thermodynamic parameter, can easily convert the state of Li-Cs from strongly phase separating to strongly long-range ordering. Due to increasing charge transfer from Cs to Li at high pressures, Zhang et al.[21] predicted that the stable phases in Li-Cs mixture are LiCs at 160 GPa, and $Li_7Cs$ at 80 and 160 GPa within the density-functional theory (DFT). Subsequent experimental study by in situ synchrotron powder x-ray diffraction[22] demonstrated that the Li-Cs alloy could be synthesized at very low pressure (>0.1 GPa), and analysis of the valence charge density also showed that electrons are donated from Cs to Li, resulting in a charge state of -1 for Li. Interestingly, Cs can also obtain electrons from Li and become anionic with a formal charge much beyond -1 at high pressures, as Botana et al[23] reported in the stable $Li_nCs$ (n=2-5) compounds under pressures above 100 GPa using first principles method within DFT scheme. This phenomenon can partially explained by tracking the variation of electronegativity between Li and Cs with pressure[24]. At 0 GPa, the electronegativity of Li (3.17) is much higher than that in Cs (1.76). Whereas at 200 GPa, the case is on the contrary (1.22 in Li lower than 1.59 in Cs).

[*] Project supported by the National Natural Science Foundation of China under (Grant Nos. 11672274, 11274281, and 11174214), the CAEP Research Projects (Grant Nos. 2012A0101001 and 2015B0101005), the Joint Fund of the National Natural Science Foundation of China and the China Academy of Engineering Physics (NSAF) (Grant No. U1430117), and the Fund of National Key Laboratory of Shock Wave and Detonation Physics of China (Grant under No. 6142A03010101).
[†] Corresponding author. E-mail: s102genghy@caep.cn;
[†] Corresponding author. E-mail: xrchen@scu.edu.cn;
[†] Corresponding author. E-mail: wuqianglsd@163.com.

However, all of these interalkalies are metallic even there is a strong charge transfer. A spectacular behavior of pure alkalies is that under compression some of them can form electride and become insulator[15, 16]. Thus one might wonder whether such an intriguing phase can occur in other interalkalies or not. Among the Li interalkalies, Li and Na have similar ionic radius[20] and the size mismatch between them is the smallest. On the other hand, they have similar electronegativity[24], which leads to the positive formation enthalpies in Li-Na[21]. Taking into account both the effects, the extent of Li-Na immiscibility is still considered to be the least at ambient pressure in comparison with those in Li-K[25], Li-Rb[26], and Li-Cs[27] system. The phase separation curve observed experimentally in Li-Na mixture showed a consolute point at 576 ± 2 K and composition $X_{Li}$=0.64[28]. Both classical molecular dynamics (CMD)[29-31] and ab initio molecular dynamics (AIMD)[32, 33] simulate satisfactorily the concentration-concentration structure factor in good agreement with the experiment data[28]. In addition, the AIMD calculations[33] suggested that the electronic density of states of the $Na_{0.5}Li_{0.5}$ alloy at the Fermi level decreases with pressures at 1000 K and a dip near the Fermi level starts to develop at high pressure of 144 GPa, which indicates a possible energy gap as observed in pure Li[34]. However, up to now, there is no theoretical or experimental evidence that Li and Na can mix to form a compound in the solid state.

In this paper, we systematically investigate the stable crystalline phases in $Li_mNa_n$ (m=1, n=1-5 and n=1, m=2-5). It is found that the unmixable Li and Na at low pressures become mixable, and LiNa is the only stable compound in Li-Na mixture at high pressures. The formation pressure in LiNa (at 355 GPa) is substantially higher than that in LiCs, in other words, the volume of LiNa is much smaller than that in LiCs, which means that the core-core overlapping between atoms in LiNa is larger than that in LiCs. This will make it more easily to transfer electrons from Li and Na atoms into the interstitial sites of LiNa, whereas LiCs has enough space to make electrons transfer between Li and Cs atoms, and is more difficult to form interstitial electrons. Remarkably, the structure of LiNa with orthorhombic oP8 symmetry is similar to that of pure Na[16], which also contains interstitial electrons that makes the material insulating.

## 2. Computational Details

Our structural prediction approach is based on a global minimization of *ab initio* total-energy calculations as implemented in the CALYPSO (Crystal structure AnaLYsis by Particle Swarm Optimization) code[35, 36], which has been successfully applied to the prediction of high-pressure structures of many systems[15, 16, 37-41]. *Ab initio* electronic structure calculations and structural relaxations are carried out by using the Vienna ab-initio simulation package (VASP)[42] with the Perdew-Burke-Ernzerhof (PBE) generalized gradient approximation (GGA) functional[43]. $1s^22s^1$ of Li and $2s^22p^63s^1$ of Na are treated as valence electrons for projected-augmented-wave pseudopotentials. The cut-off energy for the expansion of wavefunctions into plane waves is set to 650-900 eV in all calculations, and the Monkhorst-Pack grid with a maximum spacing of 0.03 Å$^{-1}$ is individually adjusted in reciprocal space to the size of each computational cell, which usually give total energies converged to ~1 meV per atom. Lattice dynamics is calculated by the small displacement method as implemented in the PHONOPY package[44].

The formation enthalpy ($\Delta H$) of Li$_m$Na$_n$ with respective to elemental Li and Na is calculated by using the following formula:

$$\Delta H\left(\mathrm{Li}_m\mathrm{Na}_n\right) = [H\left(\mathrm{Li}_m\mathrm{Na}_n\right) - mH(\mathrm{Li}) - nH(\mathrm{Na})]/(m+n) \quad (1)$$

where *H* is the enthalpy of the most stable structure of certain compositions at the given pressure. For elemental Li, the *cmca*-24 (80 GPa-185 GPa), *cmca*-56 (185 GPa-269 GPa) and *p4$_2$mbc* (>269 GPa) structures[15] are considered, and for Na, the fcc (65 GPa-103 GPa), op8 (103 GPa-260 GPa) and *hp*4 (>260 GPa) phases[16] are used.

## 3. Results and Discussion

It is known that the crystal structure is the basis for the deep understanding of any physical properties. To explore the possibility of stable Li-Na compounds, we first perform systematic crystal-structure prediction to determine the lowest-enthalpy structures of LiNa at a pressure range of 100-400 GPa. Our structure searches show that the most energetically favorable structure of LiNa adopts the orthorhombic oP8 symmetry (space group *Pnma*, 4 formula units per cell, see the inset of Fig. 1) throughout the whole pressure range considered here. Interestingly, this structure shares the same symmetry and the positions of Na atoms to the oP8 phase of elemental Na[16]. At 400 GPa, the Na atoms occupy Wyckoff site 4c (0.481, 0.750, 0.177) and the Li atoms locate at 4c (-0.156, 0.250, 0.081). Despite of the structural similarity between Na and LiNa, it seems impossible for Li and Na to form substitutional Li-Na alloy. In fact, we observed a large increase of formation enthalpy (>24 meV/atom at 400 GPa) when swapping or substituting Li and Na atoms. As indicated by our structure searches, the second-lowest-enthalpy structure of LiNa has a much larger (*e.g.*, 81 meV/atom at 400 GPa) enthalpy than the lowest one, suggesting that the local minima in the potential energy surface are deep and the LiNa-op8 can remain significantly steady. Figure 1(a) shows the calculated formation enthalpy of LiNa-op8 as a function of pressure, where it is shown that LiNa becomes thermodynamically stable at approximately 355 GPa. The lattice dynamics calculations indicate that LiNa-op8 keeps dynamically stable even when pressure decompresses down to 70 GPa, evidencing by the absence of any imaginary frequency in the whole Brillouin zone of the phonon dispersions (Fig. 2). Moreover, we also perform the structure predictions for other compositions ($Li_mNa_n$ ($m$=1, $n$=2-5 and $n$=1, $m$=2-5)) at 400 GPa, but find no other stable compounds (Fig. 1(b)).

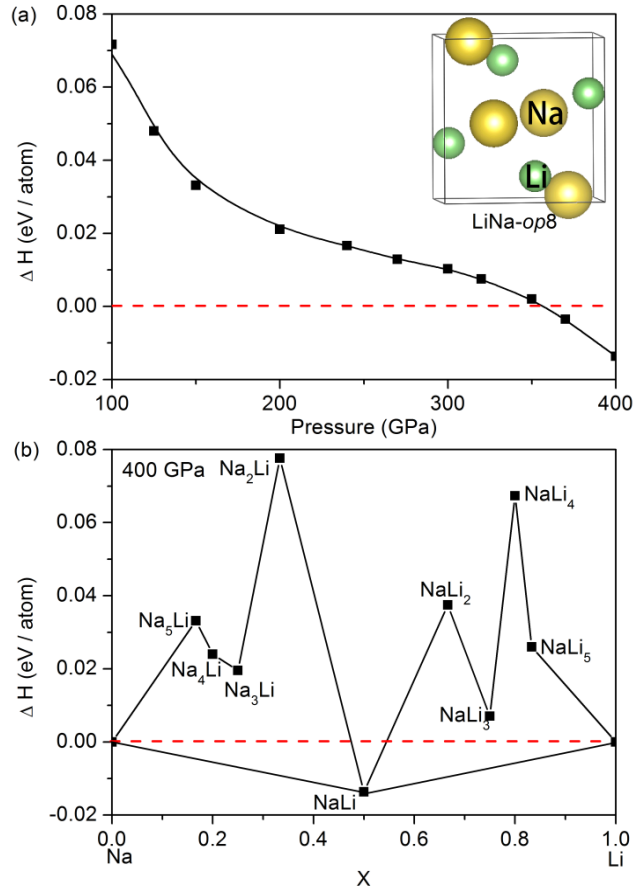

**Fig. 1.** (a) Pressure dependent formation enthalpy of LiNa calculated by PBE functional. Inset: The crystal structure of LiNa-op8; (b) Enthalpies of formation of $Li_mNa_n$ ($m$=1, $n$=1-5 and $n$=1, $m$=2-5) with respect to decomposition into elemental Li and Na at 400 GPa. The abscissa $x$ is the fraction of Na in the structures. For elemental Li, the *cmca*-24 (80 GPa-185 GPa), *cmca*-56 (185 GPa-269 GPa) and *p4₂mbc* (>269 GPa) structures[15] are considered, and for Na, the fcc (65 GPa-103 GPa), op8 (103 GPa-260 GPa) and hp4 (>260 GPa) phases[16] are used.

In the structure of LiNa-op8 at 200 GPa, the distances of neighboring Na–Na, Na-Li and Li-Li atoms are 1.99 Å, 1.80 Å and 2.65 Å, respectively. Given that the atomic radii of Na and Li are, respectively, 1.16 Å and 1.09 Å[45], it is conceivable that the core-core overlap in Na-Na and Na-Li is very strong. It is known that this overlap can cause their valence electrons repulsed by core electrons into the lattice interstices[16]. In Fig. 3(a), we plot the calculated electron density difference, in which the electron attractors locating at the lattice interstices are clearly revealed. Furthermore, we also calculate the electron-localization function (ELF), which is useful for the analysis of the degree of electron localization. As shown in Fig. 3(b), the calculated ELF with an isosurface value of 0.95 implies a high degree of localization of the electronic charge density within the voids of the crystal. This unusual high pressure phase can be viewed as a high pressure electride which is first explicitly suggested in Na-hp4 by Ma *et al*, which can be viewed as a small distortion of Na-op8 structure[16]. In these high pressure electrides, the interstitial electrons are believed to play the role of atomic anions, called the interstitial quasiatoms (ISQs)[46-48].

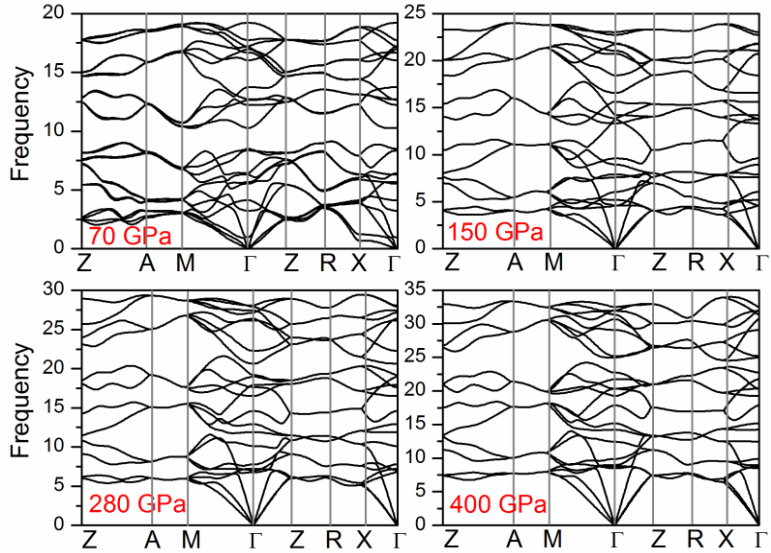

**Fig**. 2. Phonon dispersions of LiNa under different pressures.

To get further insight into the nature of electride LiNa-op8 structure, we analyzed the electron density of the ISQs with the help of Bader's effective charges[49-51]. In this method, an atom is defined as a basin that can share electron density, and a concentration of electron density in a void is attributed to ISQs. Our Bader charge analysis reveals that the ISQs are indeed negatively charged, behaving as anions. The Bader charges on Li and Na atoms are positive, indicating a charge transfer from Li and Na to the ISQs. Comparison of the Bader charges of Li and Na in electride LiNa shows that the charge of Li is a bit larger in magnitude than that of Na, e.g., at 100 GPa, the Bader charges are +0.65, +0.45 and −1.10 for Li, Na and the ISQs, respectively. This is anomalous since Li has a smaller atomic core and a resultant larger Pauling electronegativity than Na[24]. As pressure increases, the charge of ISQs also increases (+0.64, +0.53 and −1.17 for Li, Na and the ISQs at 400 GPa), indicating the increased electron localization in the voids with compression. This charge increment of ISQs originates from the charge transfer from Na rather than Li (Fig. 3(c)). Additionally, it should be noted that the Bader charge is usually somewhat smaller than the nominal ionic charge; e.g., at ambient condition, the Bader charge of Na in NaCl is only 0.78. It is thus convincible that the integrated electron densities of Na and Li in its region of stability may be approximately 1 electron, and therefore the basin of ISQ may take up 2 electrons. Note that the number of ionic cores ($Li^+$ and $Na^+$) is exactly twice that of interstitial electron density maxima. In this point of view, the structure of electride LiNa is analogous to anticotunnite-type ($PbCl_2$) structure as e·(Na, Li). Furthermore, such high ISQ charge density in electride LiNa indicates strong ionic interactions between the ISQs and the Li/Na atoms (Fig. 3(d)), which contributes significantly to the lattice stability of LiNa. In addition, it is found that other unstable $Li_mNa_n$ compounds also have interstitial electrons (in the Appendix), but their concentration is much lower than that in LiNa, which leads to weaker ionic interactions between the ISQs and the Li/Na atoms. This may be the reason that only LiNa is stable in Li-Na mixture. It is substantially different from atoms share or exchange electrons in common compounds and alloys.

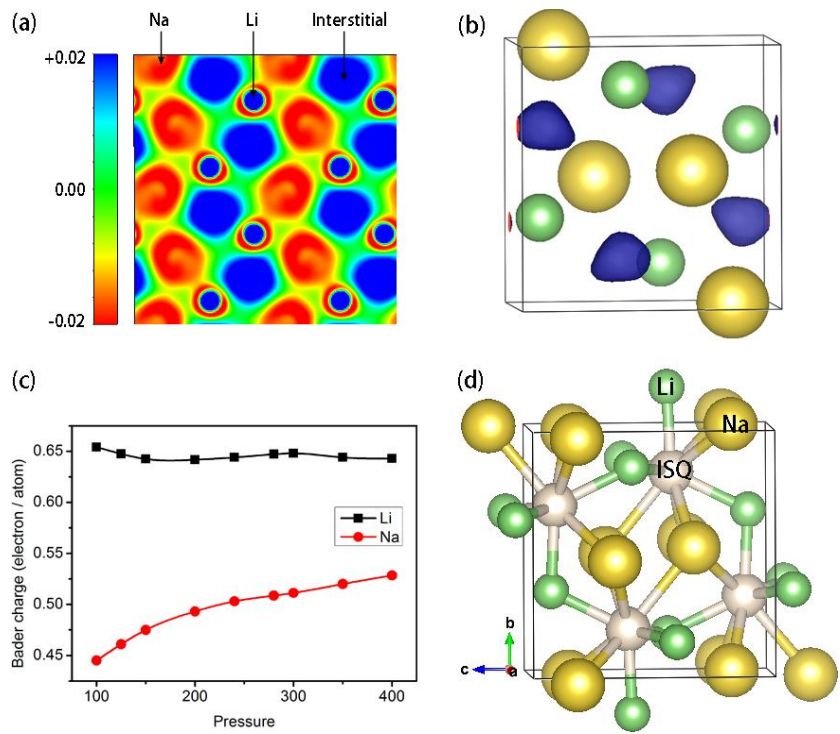

**Fig. 3.** (a) Difference charge density (eÅ$^{-3}$) of LiNa plotted in the (100) plane at 200 GPa; (b) Electron localization functions (isosurface = 0.95) of LiNa at 200 GPa. The blue areas represent strong electron localization in the lattice interstitial. (c) Pressure dependence of Bader charge of Li and Na atoms in LiNa. (d) Crystal structure of electride LiNa, wherein the white atoms represent the interstitial quasiatoms (ISQs).

Metallization is presumed to be the general trend of all materials under sufficiently strong compression. However, pressure-induced metal-insulator transitions in elemental Na and Li as well as other materials (such as Ca, Mg and Al) have received a lot of attention recently[15, 16, 52-55]. We have calculated the electronic band structures for LiNa at different pressures, as illustrated in Figs. 4(a)-4(c). It is obvious that LiNa shows insulating nature at pressures above 100 GPa, indicating that the metal-insulator transition in LiNa occurs at 100 GPa. This opening up of the band gap enhances the stability of LiNa, as corroborated by the disappearance of softening of TA mode at $\Gamma$ point and LA mode at $X$ point when pressure increased from 70 GPa to 400 GPa as shown in Fig. 2. For comparison, the gap opens up in elemental Li from 60 to 200 GPa[15] and in elemental Na from 200 GPa to 1.55 TPa[6, 16]. The pressure dependence of band gap of LiNa is plotted in Fig. 4(d), wherein a rapid increase of the gap with pressure is observed, because of the increased electron localization with compression. It should be noted that standard DFT tends to underestimate band gaps of materials due to self-interaction errors, and this problem can be partly solved by using all-electron GW approximation[56-59]. Our GW calculations indicate that the metal-nonmetal transition pressure in LiNa is 70 GPa, and when pressure increases to 400 GPa, the band gap reaches 3.67 eV (Fig. 4(d)).

In addition, figure 5 shows the calculated total and atom-projected densities of states (DOS) at 400 GPa. Notably, the occupied states in the vicinity of Fermi level ($E_f$) primarily consist of hybridized ISQ-s, Na-s, Li-p, Na-p and Na-d states, which are similar to the cases of high-pressure insulating phases in Li, Na, and Ca. The splitting of the bonding and antibonding states due to this hybridization is the main reason for the development of the band gap and emergence strong electron localization in the crystal interstices of LiNa.

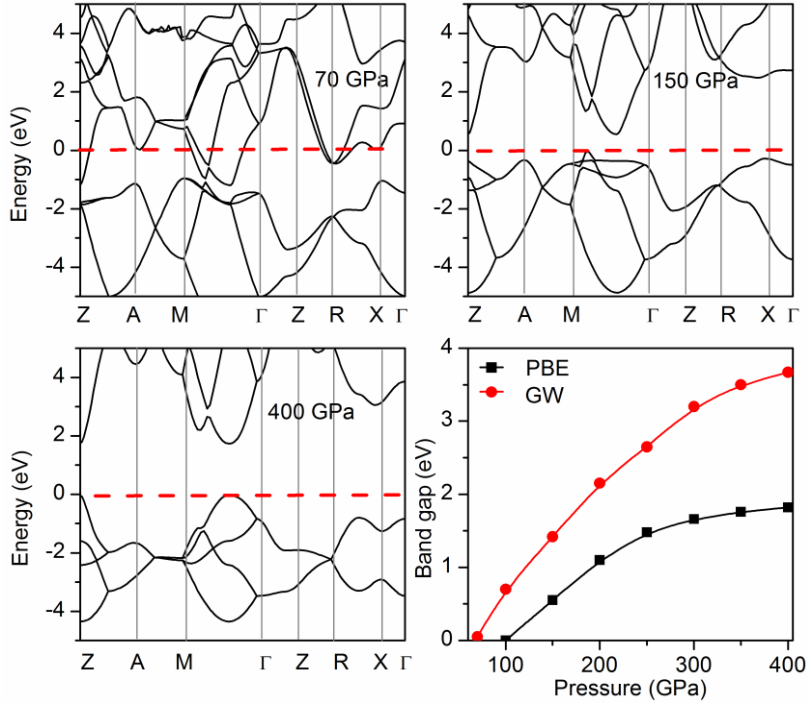

**Fig. 4.** (a-c) Electronic band structures of LiNa at selected pressures. The red dashed line denotes the Fermi energy. (d) Band gaps calculated by PBE and GW as a function of pressure.

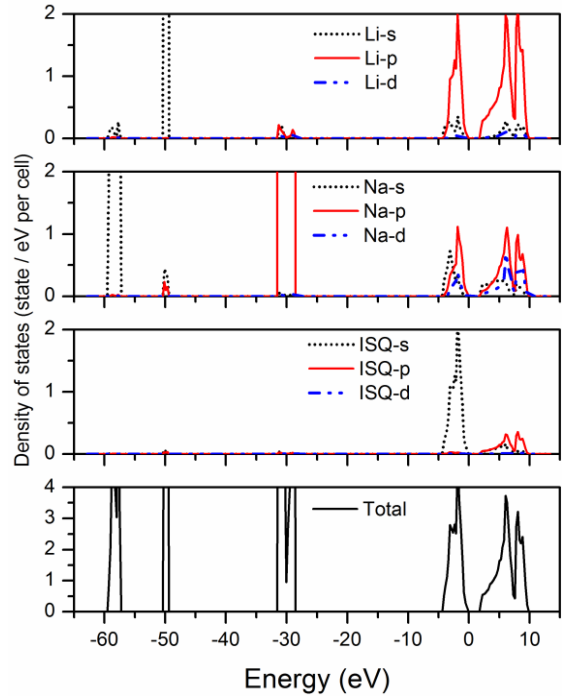

**Fig. 5.** Electronic density of states of LiNa at 400 GPa by PBE method.

## 4. Conclusions

In summary, we have explored the possibility of stable Li-Na compunds in $Li_mNa_n$ ($m$=1, $n$=1-5 and $n$=1, $m$=2-5) using the first principles calculations and a swarm structure search technique. A novel stoichiometric LiNa compound is predicted to be stable up to 400 GPa. Calculations of the electronic properties reveal that LiNa is not a metallic alloy but an insulating electride compound. Further analysis indicates that the stabilization of this compound is due to the localized interstitial electrons. Since LiNa is metastable down to 70 GPa and all other structures lie high above in energy, the insulating phase might be synthesizable in diamond-anvil cell (DAC) with thermal annealing. We believe that this study will extend the understanding of high-pressure alkali alloys and electrides.

**Appendix A：Electron localization functions of Li$_m$Na$_n$ (*m*=1, *n*=2-5 and *n*=1, *m*=2-5)**

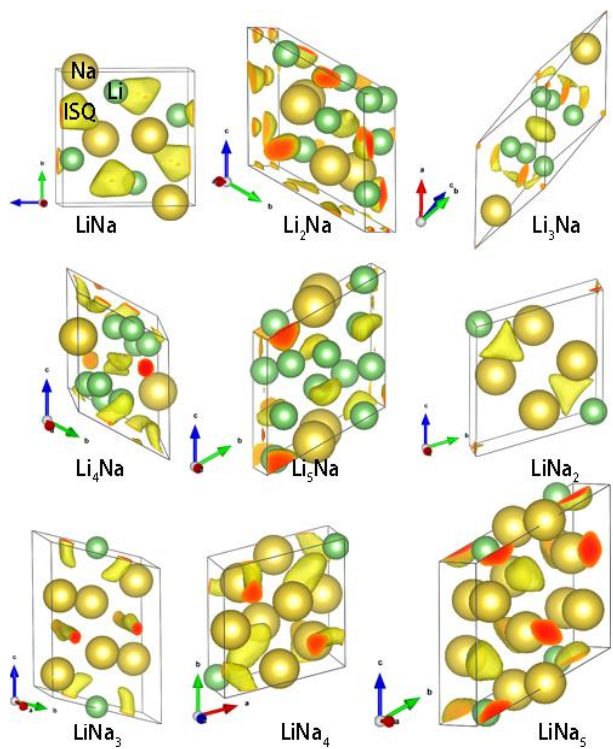

**Fig. 6.** Structures and electron localization functions (isosurface = 0.85) of Li$_m$Na$_n$ (*m*=1, *n*=2-5 and *n*=1, *m*=2-5) at 400 GPa.

## Figures

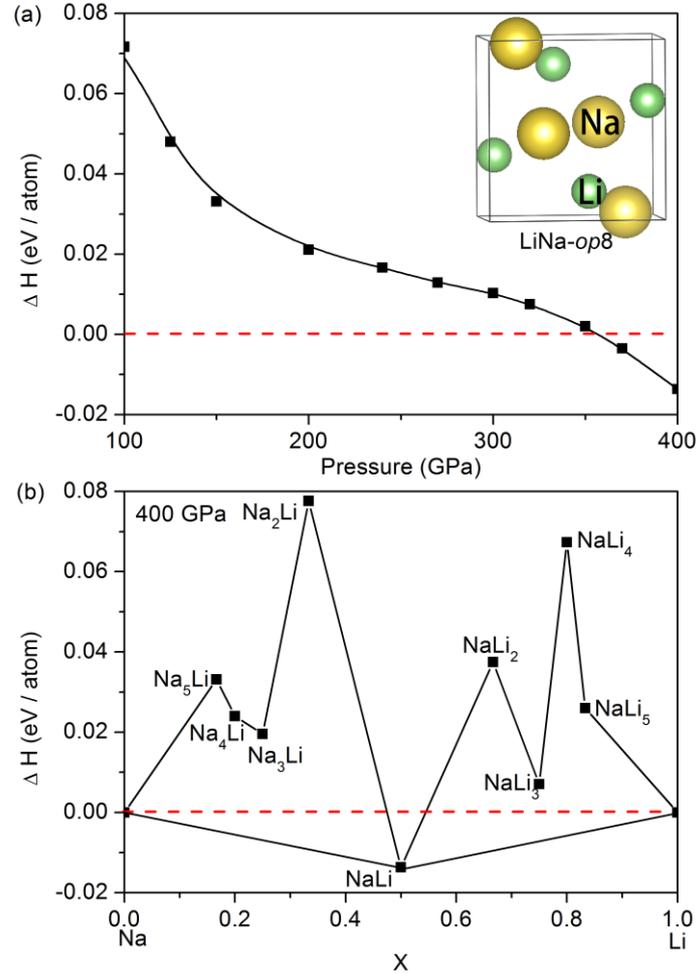

**Fig. 1.** (a) Pressure dependent formation enthalpy of LiNa calculated by PBE functional; Inset: The crystal structure of LiNa-*op*8; (b) Enthalpies of formation of Li$_m$Na$_n$ ($m$=1, $n$=2-5 and $n$=1, $m$=2-5) with respect to decomposition into elemental Li and Na at 400 GPa. The abscissa $x$ is the fraction of Na in the structures. For elemental Li, the *cmca*-24 (80 GPa-185 GPa), *cmca*-56 (185 GPa-269 GPa) and *p4$_2$mbc* (>269 GPa) structures[15] are considered, and for Na, the fcc (65 GPa-103 GPa), *op*8 (103 GPa-260 GPa) and *hp*4 (>260 GPa) phases[16] are used.

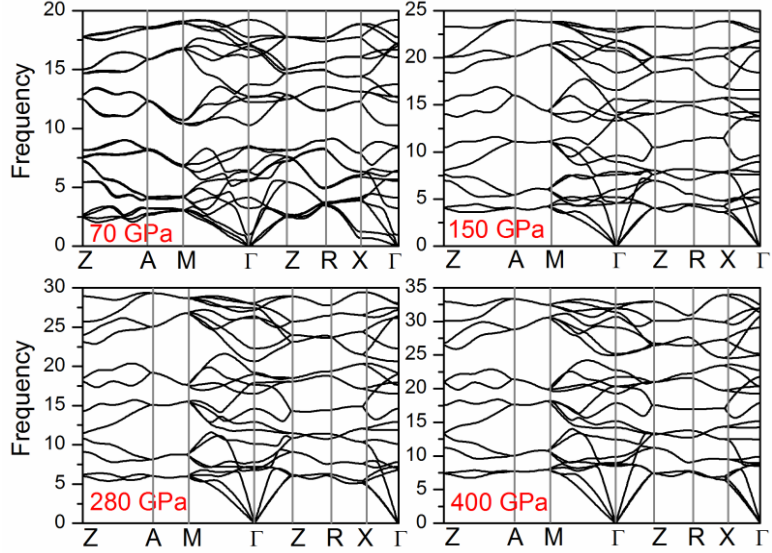

**Fig**. 2. Phonon dispersions of LiNa under different pressures.

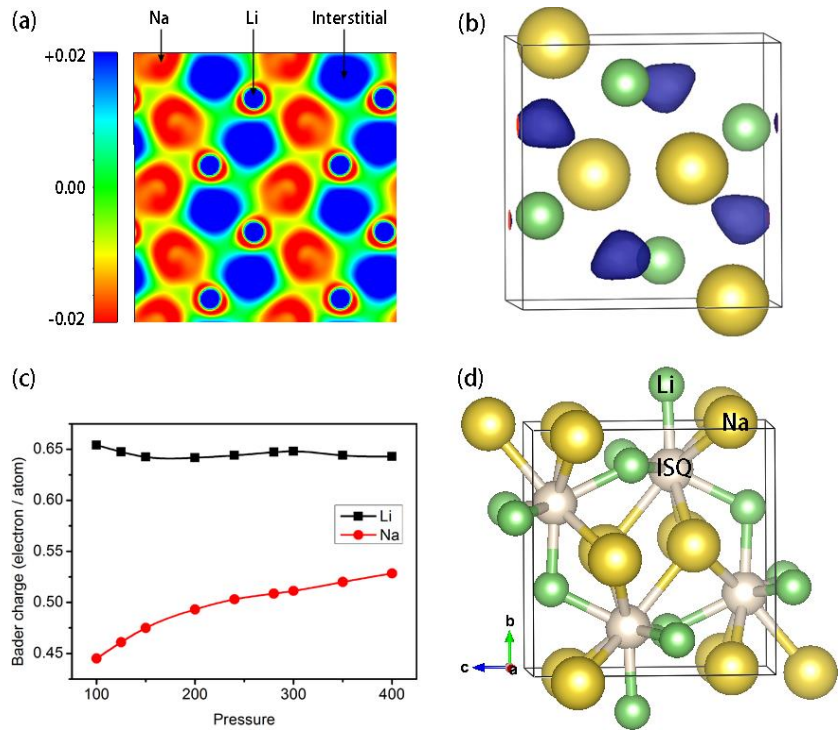

**Fig. 3.** (a) Difference charge density (eÅ$^{-3}$) of LiNa plotted in the (100) plane at 200 GPa; (b) Electron localization functions (isosurface = 0.95) of LiNa at 200 GPa. The blue areas represent strong electron localization in the lattice interstitial. (c) Pressure dependence of Bader charge of Li and Na atoms in LiNa. (d) Crystal structure of electride LiNa, wherein the white atoms represent the ISQs.

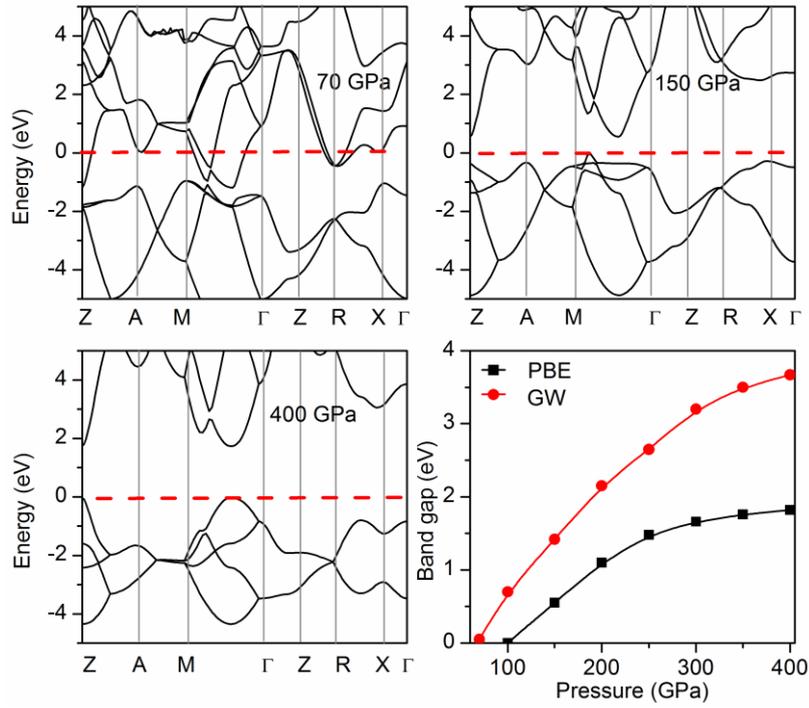

**Fig. 4.** (a-c) Electronic band structures of LiNa at selected pressures. The red dashed line denotes the Fermi energy. (d) Bandgaps calculated by PBE and GW as a function of pressure.

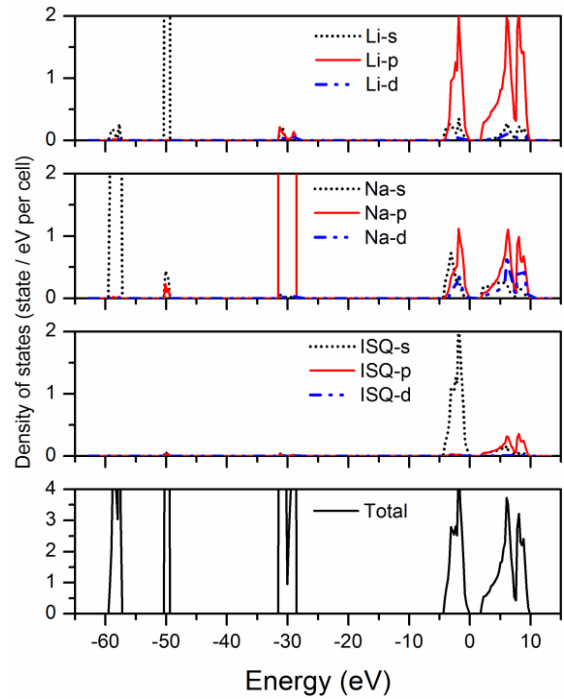

**Fig. 5.** Electronic density of states of LiNa at 400 GPa by PBE method.

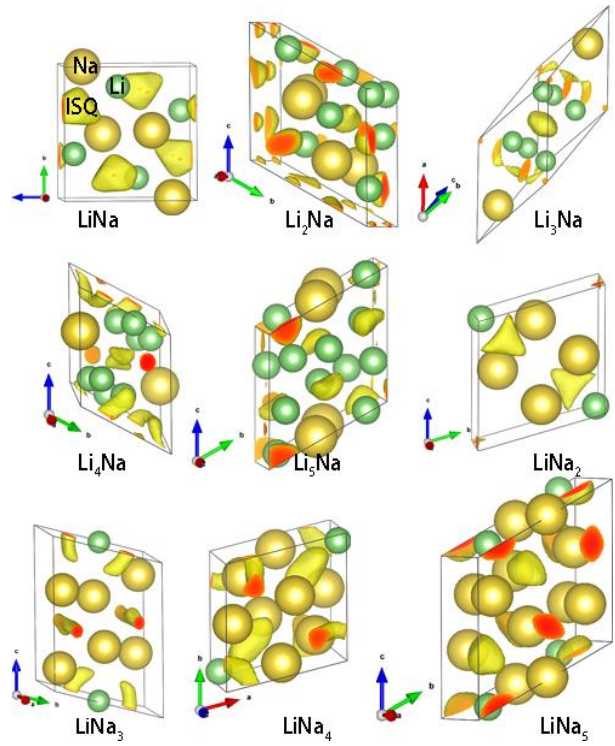

**Fig. 6.** Structures and electron localization functions (isosurface = 0.85) of $Li_mNa_n$ ($m$=1, $n$=1-5 and $n$=1, $m$=2-5) at 400 GPa.